# Two-dimensional $O(n)$ Models with Defects of "Random Local Anisotropy" Type

A.A. Berzin[a], A.I. Morosov[b]*, and A.S. Sigov[a]

[a]MIREA - Russian Technological University, 78 Vernadskiy Ave., 119454 Moscow, Russian Federation

[b]Moscow Institute of Physics and Technology (National Research University), 9 Institutskiy per., 141700 Dolgoprudny, Moscow Region, Russian Federation

* e-mail: mor-alexandr@yandex.ru

## Abstract

The phase diagram of two-dimensional systems with continuous symmetry of the vector order parameter containing defects of the "random local anisotropy" type is investigated. In the case of a weakly anisotropic distribution of the easy anisotropy axes in the space of the order parameter, with decreasing temperature, a smooth transition takes place from the paramagnetic phase with dynamic fluctuations of the order parameter to the Imri-Ma phase with its static fluctuations. In the case when the anisotropic distribution of the easy axes induces a global anisotropy of the "easy axis" type that exceeds a critical value, the system goes into the Ising class of universality, and a phase transition to the ordered state occurs in it at a finite temperature.

## 1. Introduction

In two-dimensional systems with continuous symmetry of the $n$-component vector order parameter ($O(n)$ model), a long-range order at any temperature different from the absolute zero is absent. In the case of the two-dimensional $X$-$Y$ model ($n = 2$), at finite temperature $T_{BKT}$ different from zero, a phase transition occurs from the paramagnetic phase with the exponentially decreasing correlation function of the order parameter to the Berezinsky-Kosterlitz-Thoules (BKT) phase with the power-law character of the correlation function decrease [1-3]. As shown in our recent paper [4], the violation of the $O(n)$ symmetry of the order parameter in the two-dimensional model due to the presence of an arbitrarily weak homogeneous anisotropy leads to the fact that instead of a phase transition to the BKT phase at $T = T_{BKT}$, a transition to the phase with the long-range order takes place.

A similar consideration for the two-dimensional Heisenberg model ($n = 3$) was carried out much earlier [5]. The violation of the $O(3)$ symmetry of the order parameter by a weak homogeneous anisotropy of the "easy axis" type transfers the system to the Ising class of universality and causes the appearance of the long-range order at temperature significantly lower than the transition temperature in the three-dimensional Heisenberg model, but nonzero. If the $O(3)$ symmetry is broken by the "easy plane" type anisotropy, then the Heisenberg model transforms into the universality class of the $X$-$Y$ model, and in it there is a phase transition from the paramagnetic phase to the BKT phase in the same temperature range as in the case of the "easy axis" anisotropy type.

If global anisotropy is created by the anisotropic distribution of random fields of the "random local field" type defects or by the anisotropic distribution of easy axes of anisotropy of the "random local anisotropy" type defects, it is necessary to take into account the fact that the defects also create large-scale fluctuations of the random field or random anisotropy. These fluctuations can lead to the destruction of the long-range order and the appearance of the inhomogeneous Imri-Ma phase [6], in which the order parameter follows the random field (random anisotropy) static spatial fluctuations. Thus, one should investigate under what conditions one of the two opposing trends will prevail over the other.

The case of defects of the "random local field" type was considered in our paper [4]. The present paper is devoted to the consideration of the two-dimensional systems with defects of the "random local anisotropy" type.

## 2. The energy of the classical spins system

The exchange interaction energy of $n$-component localized spins $\mathbf{s}_i$ of a fixed unit length (the length of the vector can be included in the corresponding interaction constants or fields) forming a two-dimensional square lattice in the approximation of the interaction of nearest neighbors has the form

$$W_{ex} = -\frac{1}{2}J \sum_{i,\delta} \mathbf{s}_i \mathbf{s}_{i+\delta}, \tag{1}$$

where $J$ is the exchange integral, the summation over $i$ is carried out over the entire lattice of spins, and over $\delta$ it is carried out over the nearest neighbors to given spin.

The energy of spins interaction with defects of the "random local anisotropy" type has the form

$$W_{def} = -\frac{1}{2}K_0 \sum_l (\mathbf{s}_l \mathbf{n}_l)^2, \tag{2}$$

$K_0 > 0$ is the random anisotropy constant and summation is performed over point defects randomly distributed in crystal lattice nodes, and $\mathbf{n}_l$ is the unit vector along a random easy axis.

## 3. Two-dimensional $X$-$Y$ Model with Defects of "Random Local Anisotropy" Type

We discuss anisotropic distribution of easy axes of defects in the space of the two-dimensional order parameter with the probability function of the form

$$\rho(\boldsymbol{n}) = A\big[n_x^2 + (1+\varepsilon)n_y^2\big], \tag{3}$$

where $n_x$ and $n_y$ are projections of vector $\boldsymbol{n}$ on the axes of Cartesian coordinate system, $\varepsilon > 0$ is the distribution asymmetry measure, and $A$ is the normalization constant.

As a result of averaging the random anisotropy over a given distribution, we obtain a global anisotropy of the "easy axis" type with energy per square cell [7] equal to

$$w_{an} = -\frac{\varepsilon c K_0}{4(2+\varepsilon)} s_y^2 \equiv -\frac{1}{2} K_{eff} s_y^2, \tag{4}$$

where $c$ is the dimensionless concentration of defects (their number per cell), and $K_{eff}$ is the constant of global anisotropy induced by defects. The presence of anisotropy of the "easy axis" type transfers the system to the Ising class of universality and causes the appearance of the long-range order at temperature $T_c$, which can be estimated from the condition that the temperature and anisotropy energy $E_c \sim K_{eff}\xi^2$ of a region with a radius equal to the correlation radius $\xi$ of a defect-free system [4] are equal to each other. One has [3]

$$\xi = \exp\big(b\tau^{-1/2}\big), \tag{5}$$

where $\tau = (T - T_{BKT})/T_{BKT}$ and $T_{BKT} = \pi J/2$. To the first approximation [4],

$$\tau_c \approx \frac{4b^2}{\ln^2 \frac{J}{K_{eff}}} \ll 1, \tag{6}$$

and $T_c = (1 + \tau_c)T_{BKT}$.

Formula (6) is obtained neglecting the disordering effect of random anisotropy axes. When the vector of the order parameter on the spatial scale $L$ follows the space fluctuations of the easy axis direction, there is a gain in the bulk anisotropy energy density compared to the homogeneous state. As the result, the disordered Imri-Ma phase occurs.

The optimal scale $L^*$ and the addition to the homogeneous state energy (per unit cell) $w_{I-M}$ in the two-dimensional system are [7]

$$L^* \approx \frac{J}{c^{1/2}K_0}, \qquad (7)$$

$$w_{I-M} \approx -c\frac{K_0^2}{J}. \qquad (8)$$

As suggested in our paper [4], the transition with decreasing temperature from the paramagnetic phase with dynamic fluctuations of the order parameter to the Imri-Ma phase with static fluctuations occurs at temperature $\tilde{T} = (1 + \tilde{\tau})T_{BKT}$, which is found from the condition $\xi = L^*$. In the case of defects of the "random local anisotropy" type, one obtains

$$\tilde{\tau} \approx \frac{4b^2}{\ln^2\left(\frac{J^2}{cK_0^2}\right)} \approx \frac{4b^2}{\ln^2\left(\frac{J}{K_{cr}}\right)} \ll 1. \qquad (9)$$

If the value $|w_{an}|$ for $s_y = 1$ exceeds $|w_{I-M}|$, which corresponds to the values of the global anisotropy constant $K_{eff}$ that exceed the critical value $K_{cr} \sim \frac{cK_0^2}{J}$, then $\tau_c > \tilde{\tau}$, and the transition from the paramagnetic phase to the state with the long-range order takes place. In this case, large-scale fluctuations of the order parameter do not occur; only local distortions of the order parameter near defects do occur. Otherwise, there is a transition from the paramagnetic phase to the Imri-Ma state. Since the values $K_{eff}$ and $K_{cr}$, in contrast to the case of the "random local field" type defects, are equally dependent on the concentration of defects, the parameter determining the behavior of the system is the degree of asymmetry $\varepsilon$ of the easy axes distribution. The condition of the Imri-Ma phase existence has the form

$$\varepsilon < (1 - 10)\frac{K_0}{J}. \qquad (10)$$

The "temperature - asymmetry" phase diagram of the two-dimensional X-Y model with defect-induced anisotropy is shown in Fig. 1.

## 4. Two-dimensional Heisenberg Model with "Random Local Anisotropy" Type Defects

Consider now the following form of the anisotropic distribution of the directions of random easy axes of anisotropy in the three-dimensional space of the order parameter

$$\rho(\boldsymbol{n}) = A\big[n_x^2 + n_y^2 + (1+\varepsilon)n_z^2\big]. \tag{11}$$

Averaging over this distribution gives the effective anisotropy

$$w_{an} = -\frac{\varepsilon c K_0}{5(3+\varepsilon)} s_z^2 \equiv -\frac{1}{2} K_{eff} s_z^2. \tag{12}$$

For $\varepsilon > 0$, the global anisotropy of the "easy axis" type arises in the system, which transfers the system to the class of Ising models. The temperature of the long-range order appearance in the two-dimensional Heisenberg model with weak anisotropy was obtained in Ref. [5]

$$T_c \approx \frac{4\pi J}{\ln\left(\frac{J}{|K_{eff}|}\right)}. \tag{13}$$

This expression can be obtained from the relation $T_c \sim K_{eff}\xi^2$ with the use of formula for the correlation radius in the two-dimensional Heisenberg model [3]

$$\xi = \exp\left(\frac{2\pi J}{T}\right). \tag{14}$$

In the case of the two-dimensional Heisenberg model, the condition $\xi = L^*$ gives the value of the quantity $\tilde{T}$

$$\tilde{T} \approx \frac{2\pi J}{\ln L^*} \approx \frac{4\pi J}{\ln\left(\frac{J^2}{cK_0^2}\right)} \approx \frac{4\pi J}{\ln\left(\frac{J}{K_{cr}}\right)}. \tag{15}$$

The condition for the existence of the Imri-Ma phase $\tilde{T} > T_c$ leads to a restriction on the degree of asymmetry (10). With greater asymmetry, the long-range order arises in the system, similar to that considered in the previous section. The phase diagram of the two-dimensional Heisenberg model with defect-induced anisotropy of the "easy axis" type in the variables "temperature - asymmetry" is exposed in Fig. 2.

In the case of $-1 < \varepsilon < 0$, defects induce the global anisotropy of the "easy plane" type in the system. The appearance of such anisotropy transfers the system to the class of *X-Y* models. If $\tilde{T} > T_c$, then for $T < \tilde{T}$ in the system, static fluctuations of the order parameter arise in the three-dimensional space of the order parameter (the Imri-Ma state with $n = 3$). Otherwise $(T_c > \tilde{T})$ for $T < T_c$, a transition occurs to the effectively two-component order parameter lying in the *xy* plane. To study the behavior of the system that has arisen, one needs to project random easy axes of the defects on the easy plane and use a part of the phase diagram for the *X-Y* model with defects of the "random local anisotropy" type (obtained in the previous section) corresponding to the temperature range $T < T_c$.

Since $T_c$ for the Heisenberg model with weak anisotropy is much smaller than $T_{BKT}$ for the two-dimensional *X-Y* model and the projections of the local anisotropy axes on the *xy* plane are distributed isotropically in this subspace of the order parameter, the transition to two-dimensionality is accompanied by a transition to the Imri-Ma phase with static fluctuations of the order parameter in the two-dimensional subspace of the order parameter (the Imri-Ma state with $n = 2$).

The phase diagram of the two-dimensional Heisenberg model with defects-induced anisotropy of the "easy plane" type in the variables "temperature - asymmetry" is shown in Fig. 3.

## 5. Conclusions

Finally, we formulate the main conclusions of the work:

1. The presence of defects of the "random local anisotropy" type in the two-dimensional *X-Y* model in the case of an isotropic or weakly anisotropic easy axes distribution in the order parameter space leads to the transition to the disordered Imri-Ma state at temperature close to the temperature of the transition to the Berezinsky-Kosterlitz-Thoules phase in the system free of defects. With greater asymmetry of the easy axes distribution at the same temperature, the phase transition to the state with the long-range order takes place.

2. In the two-dimensional Heisenberg model with defects of the "random local anisotropy" type in the case when the defects induce global anisotropy of the "easy axis" type, the situation is similar to the case of the *X-Y* model, except that the temperatures of the transition to the Imri-Ma state and to the ferromagnetic phase are small compared to the transition temperature in the three-dimensional system and depend logarithmically on the concentration of defects.

3. If defects induce anisotropy of the "easy plane" type, then with decreasing temperature, the transition to the Imri-Ma phase occurs with static two-dimensional (strong asymmetry) or three-dimensional (weak asymmetry) fluctuations of the order parameter.

**Financing of the work**

The work was supported by the Ministry of Science and Higher education (State assignment, project № 8.1183.2017PCh).

**Figure Captions**

1. Phase diagram "temperature – asymmetry" of the two-dimensional *X-Y* model with the anisotropy induced by the "random local anisotropy" type defects, for $K_0/J = 10^{-2}$ and $c=10^{-2}$: *P* denotes the paramagnetic phase, *F* denotes the ferromagnetic phase, and *I-M* denotes the disordered Imry-Ma phase.

2. Phase diagram "temperature – asymmetry" of the two-dimensional Heisenberg model with the "easy axis" type anisotropy induced by the "random local anisotropy" type defects, for $K_0/J = 10^{-2}$ and $c=10^{-2}$: *P* denotes the paramagnetic phase, *F* denotes the ferromagnetic phase, and *I-M* denotes the Imry-Ma phase.

3. Phase diagram "temperature – asymmetry" of the two-dimensional Heisenberg model with the "easy plane" type anisotropy induced by the "random local anisotropy" type defects, for $K_0/J = 10^{-2}$ and $c=10^{-2}$: *P* denotes the paramagnetic phase and *I-M 2d*, *I-M 3d* denote the Imry-Ma phases with two- and three-dimensional fluctuations of the order parameter respectively.

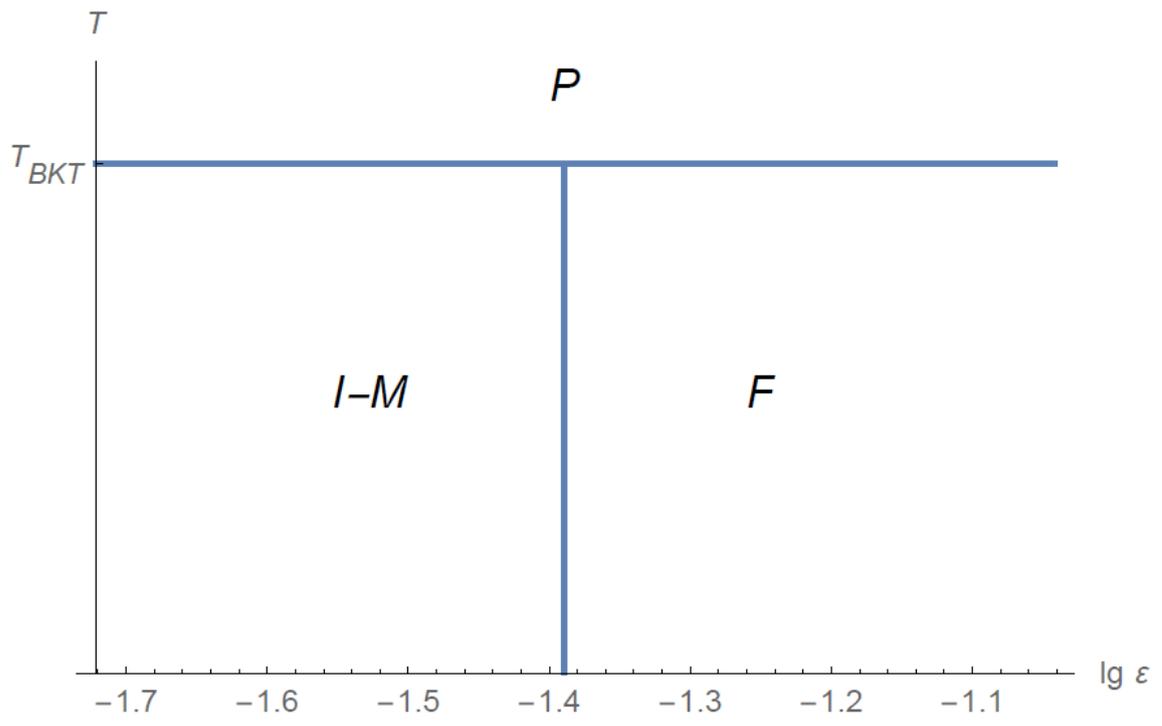

Fig. 1

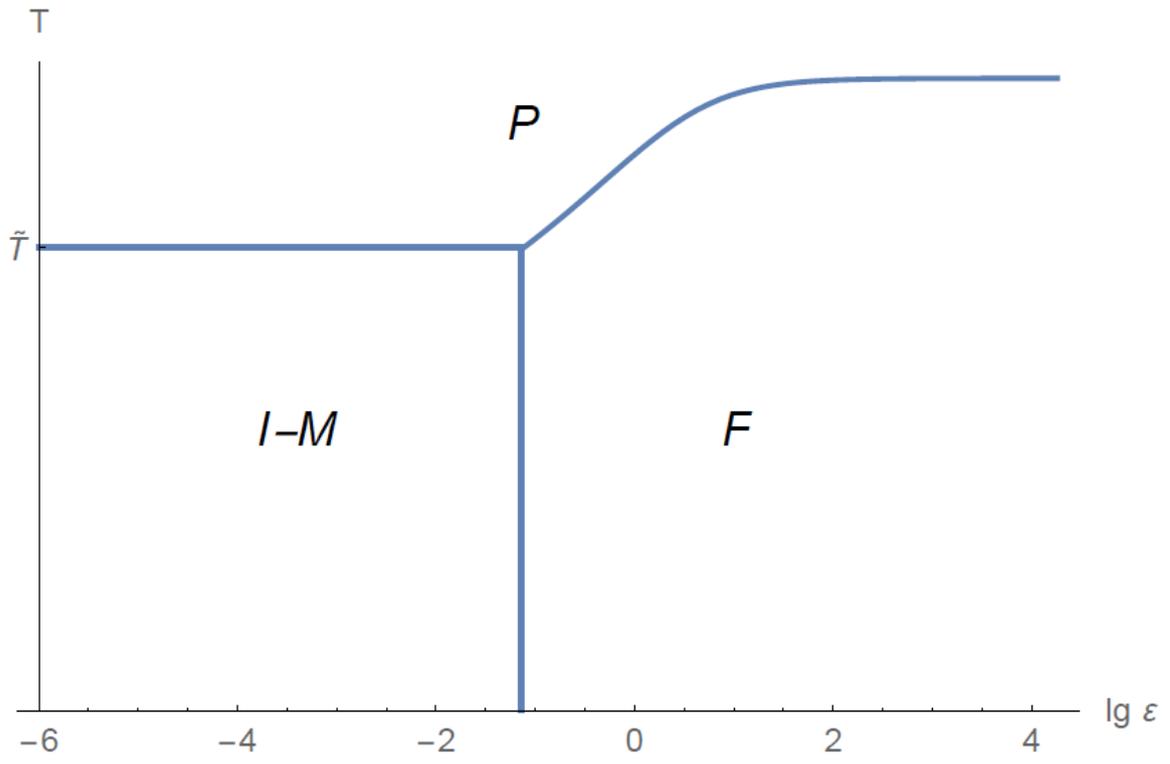

Fig. 2.

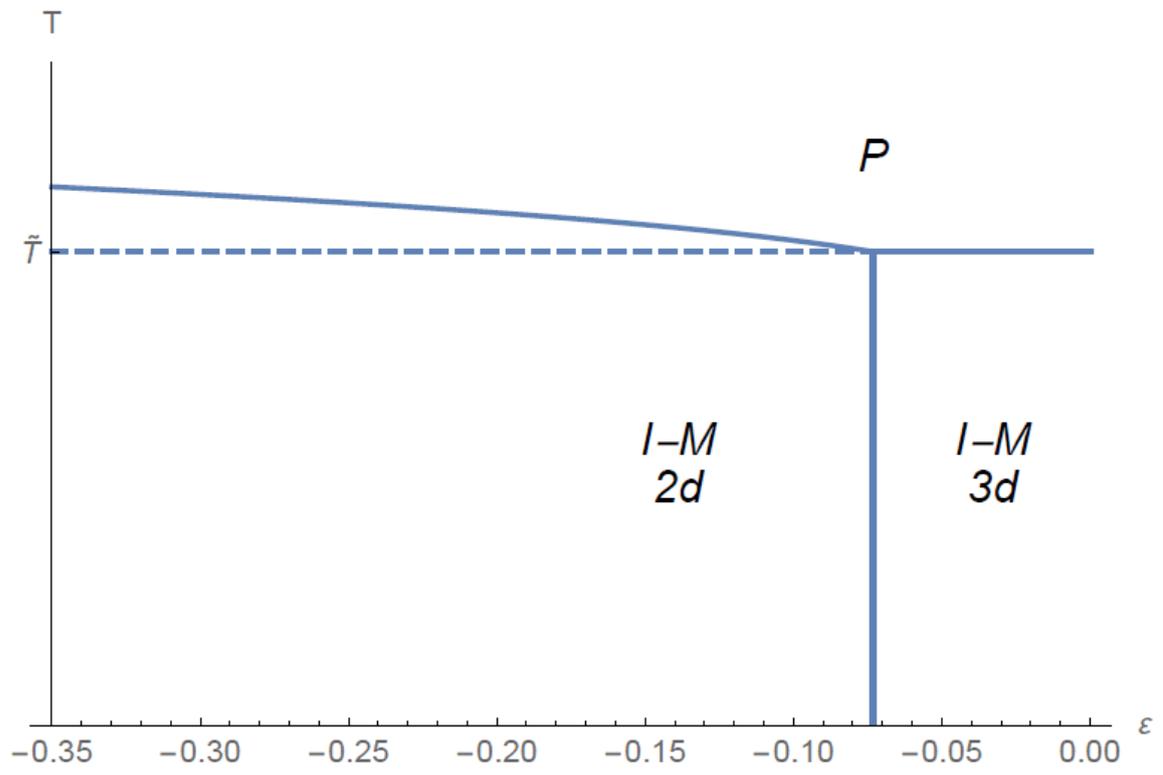

Fig. 3.